\documentstyle[aps,prl]{revtex}
\begin{document}
\draft
\twocolumn[\hsize\textwidth\columnwidth\hsize\csname @twocolumnfalse\endcsname
\title{Low Temperature Anomaly in Mesoscopic Kondo Wires}
\author{P. Mohanty$^1$
        and R. A. Webb$^2$}
\address{$^1$Condensed Matter Physics 114-36, California Institute of Technology, 
	Pasadena, CA 91125 \\
        $^2$Center for Superconductivity Research, University of Maryland, College Park, 
	MD 20742}
\maketitle

\begin{abstract}
We report the observation of an anomalous magnetoresistance in extremely dilute 
quasi-one-dimensional AuFe wires at low temperatures, along with a hysteretic
background at low fields. The Kondo resistivity does not show the unitarity limit
down to the lowest temperature, implying uncompensated spin states.
We suggest that the anomalous magnetoresistance may be understood as 
the interference correction from the accumulation of geometric phase in the 
conduction electron wave function around the localized impurity spin. 
\end{abstract}
]

 \makeatletter
\global\@specialpagefalse
\def\@oddhead{{Phys. Rev. Lett. {\bf 84}, 4481 (2000)}\hfill}
\let\@evenhead\@oddhead
\makeatother      

\par

A localized magnetic moment interacts with the conduction electrons in a metal
resulting in a logarithmic increase of the resistivity as the temperature is 
lowered. This is known as the Kondo effect \cite{kondo:book}. Below the Kondo 
temperature, $T_K$, an electron cloud begins to screen the impurity until its
spin is completely compensated, forming a singlet state at low temperature. The
nature of this state and the extent of the screening cloud has been studied for 
decades. Recently this effect has been explored in mesoscopic systems in an 
attempt to understand whether the screening is affected by the finite sample 
size \cite{giordano95,giordano96,venkat}, including high temperature large 
concentration experiments on layered Kondo systems \cite{bergmann88}, 
and 2D films \cite{haesendonck,giordano96}. Interference effects in mesoscopic 
Kondo systems containing impurity concentrations $ c > 50 \mbox{ ppm}$ do not
generally contribute significantly  to the measured magnetoresistance or resistivity
because of the strong suppression of long range phase coherence due to spin-flip
scattering. In spite of its relevance to mesoscopic systems, a complete study
of the low temperature magnetoresistance in very dilute alloys ($c < 10 \mbox{ ppm}$),
where the Kondo screening length is comparable to the phase coherence length 
$L_\phi$, has not been done. In this regime, an interference experiment 
which can reveal new information on the development of the Kondo 
screening cloud is possible.  The three-dimensional character of the local dipolar 
magnetic field from the impurity spin coupled with an externally applied field should 
provide an additional interference contribution to the electron wavefunction. This is
analogus to the Berry phase effect predicted for coherent electrons in a ring \cite{loss}
traversing in an externally applied 3D magnetic field texture concentric with the ring.

In this paper, we report the magnetoresistance and the temperature dependence of the
resistivity down to 38 mK for five quasi-1D AuFe wires in the concentration range of 
$ 3 < c < 10$ ppm. We determine both the spin-flip scattering rate and the 
phase decoherence rate by fitting the low field magnetoresistance to standard 
weak localization theory \cite{altshuler}. We find that the unitarity limit corresponding 
to the formation of the singlet state is not yet reached at our lowest temperature
\cite{altshuler2} in spite of the fact that AuFe Kondo systems are known to have
a Kondo temperature of 1 K\cite{bergmann:aufe,daybell}. At intermediate fields we observe a negative 
magnetoresistance that is characteristic in temperature dependence and shape of an 
interference correction, and different from the expected standard Kondo magnetoresistance.
At low temperatures this magnetoresistance shows hysteresis which vanishes if the magnetic
field is swept to a larger value or if the temperature is increased. We argue that
our data is not consistent with a spin glass model but rather with
a new interference correction  similar to a Berry phase effect \cite{loss}.

\par
 
Our studies are done on pure (99.9995$\%$) samples of gold (Au) before, and after, the 
ion implantation of 
3 to 10 ppm of iron (Fe) impurities. This provides a clear advantage over earlier works 
on layered or flash-evaporated samples in that the contribution to the magnetoresistance 
at various field scales coming solely from the magnetic impurities could be easily  
identified. Sample dimensions, diffusion constant 
$D$,  and $L_\phi$ measured after implantation are given in Table 1. These samples are 
quasi-1D, since $w,t \ll L_T, L_\phi$, where $L_T = \sqrt{\hbar D/k_BT}$ is the thermal 
diffusion length. The Kondo contribution to the resistivity 
$\Delta\rho(T)$ is found to have the expected logarithmic increase\cite{kondo:book}: 
$\Delta\rho (T)=A-B\ln(T)$ (See Fig.~1), after the subtraction of the electron-electron 
interaction(EEI) contribution\cite{venkat} measured before the ion implantation, 
which has the expected theoretical value\cite{altshuler}, 
$\Delta\rho_{ee} \simeq (2e^2 R^2 wt/hL^2)L_T$. 
\begin{figure}
 \vbox to 5cm {\vss\hbox to 7cm
 {\hss\
   {\includegraphics{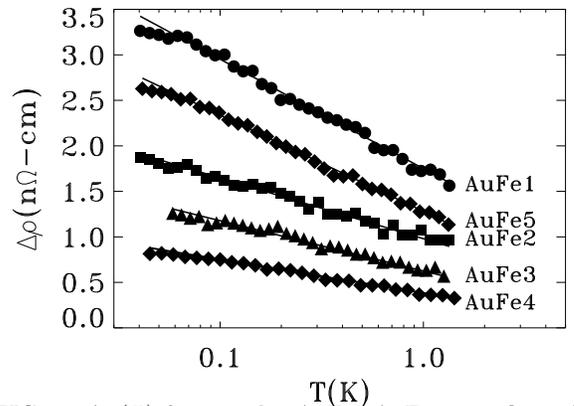}
   }
  \hss}
 }
\caption{$\Delta\rho (T)$ for samples AuFe1-AuFe5 at a finite
field. The solid lines are fits to $\ln T$. 
}
\end{figure}
\begin{figure}
 \vbox to 5cm {\vss\hbox to 7cm
 {\hss\
   {\includegraphics{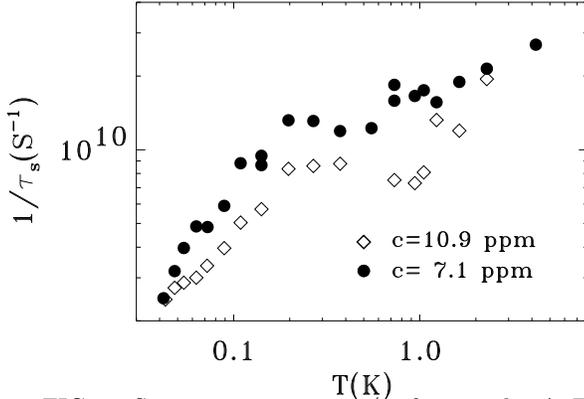}
   }
  \hss}
 }
\caption{Spin scattering rate $1/\tau_s$ for
samples AuFe1 (diamonds) and AuFe2(filled circles).
}
\end{figure}

Total scattering rate $1/\tau$ relevant for resistance is 
$1/\tau =1/\tau_{n} + 1/\tau_s$; $1/\tau_n$ is the nonmagnetic
scattering 
rate. Phase-breaking rate $1/\tau_\phi$ in the presence of magnetic scattering is given by $\tau_\phi^{-1}=2\tau_s^{-1}+\tau_{\phi(nonmag)}^{-1}$. 
Fig.~2 displays the temperature dependence of the magnetic 
scattering rate $1/\tau_s$ obtained from WL measurements\cite{bergmann88,haesendonck} 
for samples AuFe1 and AuFe2. $1/\tau_s$ is obtained from WL after subtracting 
the inelastic rate $1/\tau_{\phi(nonmag)}$ due to nonmagnetic 
sources, measured in the same Au wires before ion implantation. 
The $1/\tau_\phi$ correction term does not produce the observed behavior
seen in Fig.~2 because $1/\tau_s$ is much larger than $1/\tau_\phi$ in the corresponding
clean system. The maxima near 0.2 K-0.4 K represent the previously observed resonant
spin-flip scattering processes \cite{haesendonck,frossati}.

As shown in Fig.~1, the unitarity limit is not reached 
down to 40 mK, even in the presence of disorder and a finite magnetic 
field required to quench WL, both of which should help form the singlet state. This is
consistent with earlier observations \cite{venkat,daybell}.
The impurity spin is thus not completely
screened. However, at a larger magnetic field, a resistivity plateau is 
observed corresponding perhaps to the unitarity limit (See Fig.~3(a)). The plateau shifted to 
higher temperatures with increasing magnetic field.  Additionally, we observed a maximum 
around $T_K$ (See Fig.~3(b)). 
This observation is consistent with earlier experiments 
on (LaCe)Al$_2$ and (LaCe)B$_6$ \cite{felsch}, consequently explained by a wave 
description of the spin density \cite{weber}. This implies that there is a substantial 
spin polarization around the impurities 
with a potential $V(r) = V_0 \cos (2 k_F r)/r^3$. The local  
magnetic field 
\begin{table}{Table 1. Sample parameters shown in Figs.~1-5.}
\begin{tabular}{lcccccccccc}
Sample & w  &
t& L & $R$ & $D$ & $L_\phi$ & c & B \\ 
 & (nm) & (nm)& ($\mu m$) &$(\Omega)$ & $m^2/s$ & $(\mu m)$  & (ppm) & \\ \tableline 
AuFe1  & 180   & 40 &  155  & 393 & 0.02 & 1.9   & 10.9 & 0.52 \\ 
AuFe2  & 120   & 40 &  155  & 599 & 0.02 & 2.2   & 7.1  & 0.29 \\
AuFe3  & 100   & 35 &  155  & 803 & 0.02 & 1.7   & 6.0  & 0.24 \\  
AuFe4  & 210   & 135 & 4120  & 783 & 0.07 & 5.0   & 3.3  & 0.16 \\
AuFe5  & 120   & 135 & 2750  & 1300 & 0.05 & 3.0   & 10.1 & 0.46   \\
\end{tabular}
\end{table}
\begin{figure}
 \vbox to 6cm {\vss\hbox to 7cm
 {\hss\
   {\includegraphics{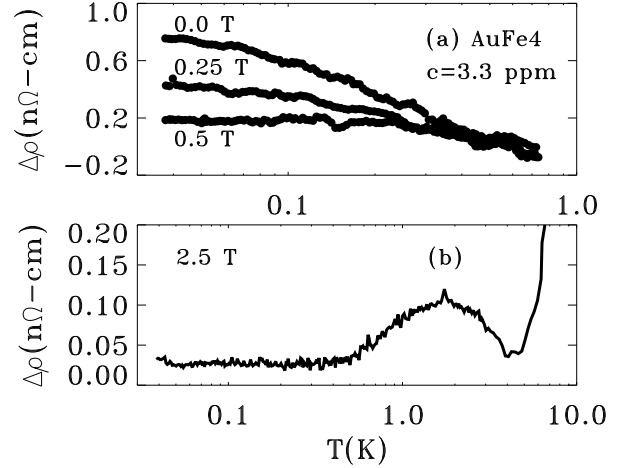}
   }
  \hss}
 }
\caption{(a) Resistivity saturation, and  
(b) the maximum in the Kondo resistivity at 2.5 T.
}
\end{figure}
\noindent
of the spin polarization can be on the order of a Tesla within a 
couple of nanometers from the impurity, though it is negligible on the scale of 
the typical inter-impurity distance of $\sim$ 10 nm. Strength of this potential $V_0$ is 
experimentally known to be very large for AuFe, decreasing exponentially with increasing 
concentration $c$ \cite{liu}. Thus, there are strong local magnetic fields for purer 
samples with longer $L_\phi$. NMR measurements of the conduction-electron 
spin density around Fe atoms in a Cu matrix also find a nonvanishing radial component
above and 
below $T_K$ \cite{slichter}.

That there exists a distribution of local magnetic fields from the impurity spins is
further confirmed by the observation of  hysteresis in the low-field MR. 
As shown in Fig.~4, the background of the WL curve is asymmetric with a positive or 
negative slope depending on the field history. Hysteresis disappears at high temperatures,
typically between 0.4 K 
and 1.5 K depending on the sample. In contrast to what is observed in a spin glass, 
we find this hysteresis to be stronger for systems with longer $L_\phi$ (hence 
for lower concentration samples). Hysteresis is expected for a spin glass system below 
$T_g$; so if it were a spin glass, we would have observed stronger hysteresis for
higher concentration samples, contrary to our data.
Our experiment suggests that hysteresis arises because of different realizations 
of the three-dimensional local field distribution. As the sample gets cold, impurity
spins freeze out in random orientations, providing a particular configuration for the 
local-field distribution. This distribution is modified by a magnetic field due to
spin alignment. 
Magnetic field cycling between $\pm$ 1 Tesla removes the hysteresis and flattens 
the background of the low-field MR, while cycling between $\pm$ 0.05 Tesla does not.

All our samples are in the single-impurity regime and the logarithmic increase of 
resistivity scales with concentration. It is unlikely that these 
systems behave like a spin glass for a number of reasons: (a) In AuFe, 
spin glass behavior is not observed for 
$c \ll 100 \mbox{ ppm}$, as is well known\cite{ford}; (b) Second, 
spin glass temperature $T_g$ 
\begin{figure}
 \vbox to 5.5cm {\vss\hbox to 7cm
 {\hss\
   {\includegraphics{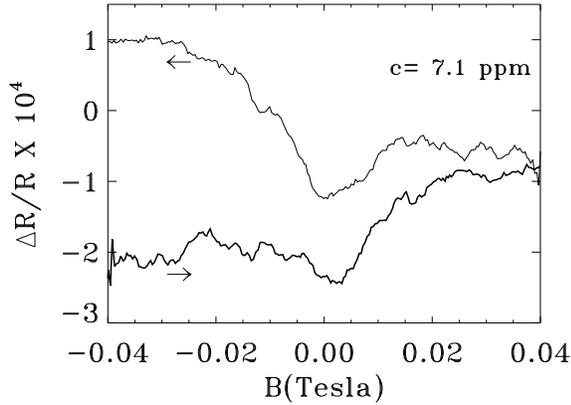}
   }
  \hss}
 }
\caption{Typical hysteresis observed in AuFe2 at 140 mK. Field sweep directions
are marked.}
\end{figure}
\noindent
for a system with 3-10 ppm Fe in Au would be 1 mK or lower; 
(c) The resistivity maximum expected for a spin glass is also not seen 
in Fig.~1; and, (d) The possibility of inhomogeneous pockets of impurities, 
or clustering, is ruled out by measuring different segments in a sample. 
The observed behavior is found to be independent of the choice of the segment, 
suggesting a homogeneous mechanism. For these reasons, the spin glass 
formation can be ruled out.

High-field magnetoresistance of a representative sample, AuFe4, 
is shown in Fig.~5(a). WL is observed at a field scale of 
$B < 0.03 \mbox{ T}$. At high fields, due to the cyclotron orbits of the electrons 
a classical MR is expected: $\Delta R_c/R \sim (\omega \tau_e)^2$, with $\omega \sim 
eB/m$, and $\tau_e$ being the electron mean free path. 
This classical $B^2$ dependence is displayed in Fig.~5(b), which is subtracted out 
in Fig.~5(a). At the intermediate field scale $(\sim 1 \mbox{ T})$, we observe a negative 
magnetoresistance in all our samples at $T < T_K$ that is very sensitive to temperature.  
Theoretically, in the standard Kondo model, one expects 
a negative MR due to the suppression of the spin-flip scattering by
the alignment of the spins with the field: $\Delta R_2/R \sim (gS\beta)^2 
(H/T)^2$, where $\beta$ is the Bohr magneton. The data is not described by this 
contribution, as evident in the shape of the MR at various temperatures. We have 
observed this
anomalous MR in all our samples along with the WL dip at zero field. 
At 40 mK, the conductance change, $\Delta G = \Delta R/R^2$, for all our samples in units 
of $e^2/h$ is : $\sim 0.001, \sim 0.002, 0.018, 0.028,$ and 0.004 for samples
AuFe1 through AuFe5 respectively.

Earlier experiments on higher concentration 
AuFe samples\cite{giordano96,venkat} revealed a behavior compatible 
with the standard expected form, and different from what we observe. 
Above $T_K$, the standard high-field magnetoresistance is essentially 
a function of the thermal average of the local moment in the field direction 
$\langle S_Z \rangle$. As temperature is increased, the field scale increases with 
the height of the MR decreasing,
ultimately becoming flat at a very 
high temperature due to thermal fluctuations of the localized spin. This behavior 
is observed in 2D 
\begin{figure}
 \vbox to 10cm {\vss\hbox to 7cm
 {\hss\
   {\includegraphics{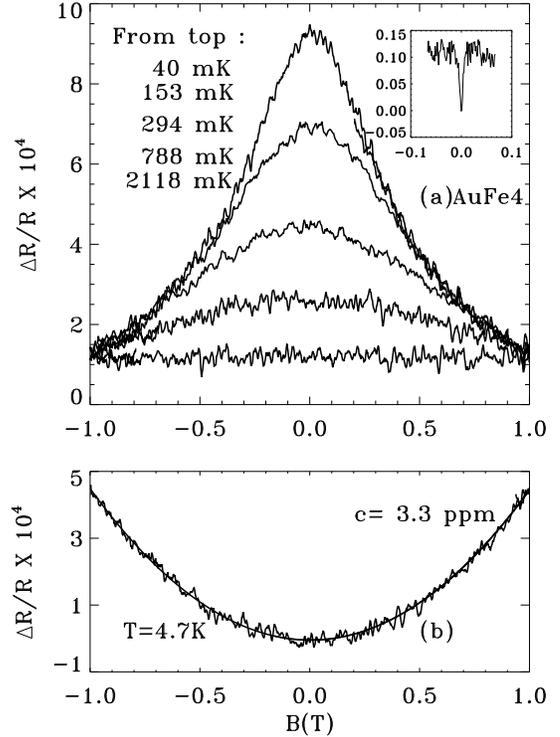}
   }
  \hss}
 }
\caption{(a) Magnetoresistance of a representative sample, AuFe4, 
after the subtraction of the classical 
magnetoresistance (shown in (b)). The inset to (a) shows the 
weak anti-localization contribution at 520 mK. 
}
\end{figure}
\noindent
Kondo films of AuFe at 1.4 K and 4.2 K \cite{giordano96}. 
However, in another experiment on AuFe wires\cite{venkat} with a much higher concentration 
of Fe impurities ($\sim 50$ ppm), temperature dependence of the MR was not studied. 
There are two important characteristics of our low temperature MR, 
different from the bulk Kondo behavior. First, the magnetoresistance as 
a function of temperature cannot be explained by $\langle S_Z \rangle$, since the field scale 
is expected to grow with increasing temperature while conserving area under 
the curve. Second, $\langle S_Z \rangle$ as a function
of temperature is expected to
increase with decreasing temperature, becoming flat at low temperatures, whereas 
the dependence shown in Fig.~5(a)  displays  no saturation  down to 40 mK.
High concentrations of impurities in the earlier experiment 
on AuFe wires \cite{venkat} and high temperature range in the experiments on 
2D AuFe films \cite{giordano96} imply a very short $\tau_\phi$ 
in these systems, yielding non-mesoscopic bulk behavior.
The local magnetic field due to polarization in these high concentration samples 
is expected to be extremely weak, in contrast to our samples. 
 
\par

It is clear from our resistivity and scattering rate measurements that the long 
range polarization of the conduction electrons around the localized spin is effective at 
low temperatures for our low concentration mesoscopic systems. From our observation
of hysteresis, we believe that this polarization or the local magnetic field causes 
the anomalous high-field magnetoresistance. Furthermore, the shape of the magnetoresistance
and its temperature dependence are very much similar to what is 
expected from a quasi-1D interference effect\cite{altshuler}, which suggests a similarity
to weak localization. These effects were seen in long $L_\phi$ 
samples, implying an essential role played by the phase coherence of electrons.
Considering all this, we propose a connection of this new interference 
correction to Berry phase.

It is possible for the phase coherent mesoscopic Kondo wires to show a 
weak-localization-like magnetoresistance driven by a geometric phase 
$\Gamma = \int_{t_i}^{t_f} {\bf A}_g \cdot d{\bf R}$, similar to the
standard weak localization driven by the
Aharonov-Bohm phase \cite{loss}. 
${\bf A}_g$ is the geometric gauge potential, and ${\bf R}$ is the 
position vector describing the tip of the spin.
The spin part of the wave function of the phase coherent electron picks up a 
geometric phase as it aligns along the local magnetic field of the uncompensated spin. 
This is further helped by disorder in the sample \cite{loss}, since the electron spends more time 
around the spin than it would in a ballistic sample. The corresponding 
geometric phase is equal to half of the solid angle subtended by the area enclosed 
by the tip of the electron spin vector  due to its evolution in a closed loop. 
A complementary path, going in the opposite direction, will contribute an opposite phase
shift. Interference of two such paths around the local field results in a correction
to conductivity, analogous to the anticipated Berry phase correction in a ring structure.
There are no oscillations as in the Aharonov-Bohm effect, but just half a period in 
resistivity, because the maximum Berry phase acquired is $\pi$, half of the maximum solid 
angle of $2\pi$. An externally applied perpendicular field aligns the electron spin. 
If the spin is completely aligned along the external field, 
the solid angle subtended by the tip of the spin is zero, resulting in the
complete suppression of the Berry phase correction.

Berry phase changes sign under time-reversal. This leads to a contribution similar 
to the Cooperon propagator in WL. Correction 
to the resistance contains the disorder average of all possible loops acquiring
Berry phase. 
As temperature is increased, $L_\phi$ (which includes spin fluctuations) 
reduces greatly, thus reducing the magnetoresistance correction as seen in Fig.~5(a). 
This dependence is similar to that of WL. In the spirit of WL, a geometric length $L_B$ 
can be introduced, which is the length scale over which the net accumulated 
geometric phase is on the order of $\pi$. $L_B$ may be defined by 
$\Delta R_g/R^2 \sim
(e^2/\hbar) L_B/L$. For the data from the sample AuFe4( shown in Fig.~5(a)) at 40 mK, 
the geometric length $L_B \sim 18 \mu m$ ($L_\phi \sim 3 \mu m$ at 40 mK), implying 
that within  $L_\phi$ the acquired (disorder-averaged) 
geometric phase is on the order of $\pi L_\phi/L_B \sim \pi/6$ for this sample.

\par

To summarize, we have observed  an unusual temperature dependence of
the magnetoresistance along with hysteresis in quasi one-dimensional 
disordered Kondo systems at $T < T_K$. 
We believe that this arises from the adiabatic 
evolution of the phase coherent electron around the impurity spin, which results
in a Berry phase effect. 
We thank B. Altshuler, H. Fukuyama, D. Loss, P. Schwab, J. Schwarz, and 
A. Zawadowski for conversations. 
This work is supported by the NSF (DMR9510416) and 
the ARO (DAAG559710330). 

 
\par

\narrowtext


\end{document}